\documentclass[aps,pre,twocolumn,groupedaddress]{revtex4}
\usepackage{natbib}
\usepackage{graphicx}
\def\be{\begin{equation}}
\def\ee{\end{equation}}
\begin{document}

\title{Coherence and incoherence in extended broad band triplet interaction\footnote{Accepted in Phys. Rev. E (2008).}}
\author{G.I. de Oliveira}
\address{Departamento de F\'{\i}sica, Centro de Ci\^encias Exatas e Tecnologia
Universidade Federal do Mato Grosso do Sul, \\Caixa Postal~549, 
79070-900 Campo Grande, MS, Brasil,} 
\author{F.B. Rizzato} 
\address{Instituto de F\'{\i}sica,
Universidade Federal do Rio Grande do Sul, \\Caixa Postal~15051, 
91501-970 Porto Alegre, RS, Brasil.}


\begin{abstract}
In the present analysis we study the transition from coherent to
incoherent dynamics in a nonlinear triplet of broad band combs of waves. 
Expanding the analysis of previous works, this paper investigates what happens when the 
band of available modes is much larger than that of the initial narrower combs 
within which the nonlinear interaction is not subjected to selection 
rules involving wave momenta. Here selection rules are present and 
active, and we examine how and when coherence can be defined. 
\end{abstract}
%
\maketitle
%

\section{Introduction}

Wave triplet interactions model a vast number of cases
where nonlinear wave dynamics of physical systems can be described in
terms of three dominant modes. The interaction is seen in a variety of
situations, ranging from three wave interactions in laser-plasma and
optical systems to pulsar emission of electromagnetic radiation, including
wave interaction in fluids and in several other settings
\cite{shu86,kiv89,vi88,gra97}. 

The conservative interaction, which will be our focus here, is more easily handled when 
the interaction involves only the three pure modes of the triplet. However, a more realistic view should 
allow for a microscopic description, where each of the pure modes is replaced with a comb with many 
submodes. This has been done in a number of papers \cite{men85,men88,rob96,fri06} where several results 
have been derived along recent years. The main lesson one learns is that the dynamics can be 
correctly described in terms of three single or central modes, as long as nonlinearities are 
strong enough to lock all submodes into a  single coherent mode. For practical purposes the coherent modes 
can then be viewed as the pure modes of the triplet interaction. On the other hand, if locking is not effective, 
each of the submodes follows its own linear dynamics and coherence is lost. Random phase approximations 
can then be invoked to analyze the problem \cite{dav72}, but the concept of a pure triplet has to be abandoned.  

A recent paper \cite{fri06} shows how combs of modes can be very naturally formed in a wave system: 
the 
essential requirement, as we shall review, is that the nonlinear interaction takes place under spatially inhomogeneous 
conditions. When 
the inhomogeneity is present, wave vector matching among the interacting modes needs not to be exact since 
it includes the reciprocal vectors of the inhomogeneities. What happens then is that even if the initial conditions 
involve a small number of wave modes, in a very short time interval the initial 
modes scatter off the inhomogeneities, creating groups of many modes, the initial combs. 
Another but equivalent way to see how combs are related to inhomogeneities is to realize that in the 
interaction of tightly packed group of modes, neighboring wave 
vectors cannot be properly resolved in finite size spatial scales, a typical occurrence in experimental 
settings \cite{wil77,men05}. In this case whole groups of 
modes with similar wave vectors are altogether excited forming the combs. The interaction acquires the aspect of a 
mean field theory, where modes of one comb interact with averages taken over modes of the remaining 
combs \cite{wil77}. 

In the past, models for wave combs were based on combs with fixed number of modes. Once the combs 
were formed, submodes could evolve in time, but always preserving a prefixed total number within 
each of the combs. A recent analysis \cite{fri05} shows that combs with prefixed number of modes cannot 
actually maintain this number if the interaction takes place in a {\it homogeneous} environment. As one may 
conclude from the comments above, this happens in virtue of the fact that homogeneity is unable to create a natural 
wave vector scale which could accommodate a given finite number of 
modes. Ref. \cite{fri05} indeed shows that as the wave dynamics develops, more and more modes 
are gradually excited and included in the interaction.

This leads us to the central question of the present analysis, namely can the wave interaction in inhomogeneous 
settings be well described with combs of finite number of modes? We shall see that the answer depends on the 
time scales and the wave vector scales one is interested in. 

The plan of the paper is the following. In \S 2 we first define a convenient interaction model allowing for an 
inhomogeneous environment and explore how the model can be used to create the picture of 
interacting combs with fixed number of modes, simultaneously analyzing its inherent limitations. 
In \S 3 we examine what happens when the constraint of a constant prefixed number of modes is relaxed. 
In \S 4 we summarize our results.

\section{The model} 

The investigation starts as we consider the set of fully dimensionless 
space time equations governing the decay  of mode ``{\it 1}'' into modes ``{\it 2}'' and ``{\it
3}'':

\begin{eqnarray}
i\partial_t a_1(x,t) + i v_{g1}  \partial_x a_1(x,t) = s(x) a_2(x,t) a_3(x,t),
\label{eq1}\\ 
i\partial_t a_2(x,t) + i v_{g2} \partial_x a_2(x,t) = s(x) a_1(x,t) a_3(x,t)^\ast,
\label{eq2}\\
i\partial_t a_3(x,t) + i v_{g3} \partial_x a_3(x,t) = s(x) a_1(x,t) a_2(x,t)^\ast.
\label{eq3}
\end{eqnarray}  
Set (\ref{eq1}) - (\ref{eq3}) actually describes the slow modulational dynamics for the complex 
wave amplitudes $a_p(x,t)$ ($p=1,2,3$) of corresponding carrier waves whose frequencies and wave vectors 
are matched. The combs are thus the multitude of sideband modes forming around each of three 
high frequency carriers. $i^2 = -1$, and the real function $s(x)$ is the spatially dependent form factor 
introducing inhomogeneity in the 
problem.  Function $s(x)$ could be typically associated with inhomogeneous density distributions in plasma 
systems for instance. 

Let us first of all see how the classical picture of combs with given number of modes 
can arise from the basic set. We first need a structure for the function $s(x)$. We define it as an 
even function centered at $x=0$ and a characteristic half width $l_s$, as follows: 
\begin{equation}
s = s(x/l_s) = s(|x|/l_s); \>\>\>\>\>\>  s (|x|/l_s \gg 1) \rightarrow 0,\>\> s(0)=1,
\label{eq4}
\end{equation}
where for mathematical convenience, and with no loss of generality, we assumed a scaling that 
renders $s(0)=1$. This kind of function restricts the effective interaction region as commented in the Introduction and can be used to 
introduce the basic wave vector associated with the inhomogeneities 
of the system in the form $k_{s} \sim 1/l_{s}$. Now we write each of the waves $a_p(x,t)$ as combs of 
many modes

\begin{equation}
a_p(x,t) = \int \hat a_p(\kappa_p,t) e^{i \kappa_p x} d\kappa_p,
\label{eq5}
\end{equation}
where $\kappa_p$ denotes the wave vectors of submodes within each comb. 

Spatial Fourier analysis of set (\ref{eq1}) - (\ref{eq3}) produces the
following group of equations for the various submodes:
\begin{widetext}
\begin{eqnarray}
i \dot {\hat a}_{1}(\kappa_1) = v_{g1}\, \kappa_1 \hat a_{1}(\kappa_1) + 
\int_{\kappa_2,\kappa_3} \hat s(\kappa_1-\kappa_2-\kappa_3) \,
\hat a_2(\kappa_2) \hat a_3 (\kappa_3) d\kappa_2 \, d\kappa_3,
\label{eq6} \\
i \dot {\hat a}_{p}(\kappa_2) = v_{g2}\, \kappa_2 \hat a_{2}(\kappa_2) + 
\int_{\kappa_1,\kappa_3} \hat s(\kappa_1-\kappa_2-\kappa_3)\, 
\hat a_1(\kappa_1) \hat a_3 (\kappa_3)^\ast d\kappa_1 \, d\kappa_3,
\label{eq7} \\
i \dot {\hat a}_{3}(\kappa_3) = v_{g3}\, \kappa_3 \hat a_{3}(\kappa_3) +  
\int_{\kappa_1,\kappa_2} \hat s(\kappa_1-\kappa_2-\kappa_3)\, 
\hat a_1(\kappa_1) \hat a_2 (\kappa_2)^\ast d\kappa_1 \, d\kappa_2,
\label{eq8}
\end{eqnarray}
\end{widetext}
with 
\begin{widetext}
\begin{equation}
s(x/l_s) = \int_{-\infty}^{\infty} \hat s(\kappa_s) e^{i \, \kappa_s \, x} d\kappa_s, 
\>\>\>\>\> \hat s(\kappa_s) = {1 \over 2 \pi} \int_{-\infty}^{\infty} s(x/l_s) e^{-i \, \kappa_s \, x} dx,
\label{eq8p5}
\end{equation}
\end{widetext}
$\hat s(\kappa_s)$ also even. One thus sees from the second of Eqs. (\ref{eq8p5}) that in general, wave vector 
mismatches of magnitudes up to $|\kappa_1 - \kappa_2 - \kappa_3 |_{max} \approx \pi /l_s$ 
among the interacting submodes are allowed. If one defines a band width $\Delta$ in the form 
$-\Delta/2 < \kappa_p < \Delta/2$, one 
concludes that all modes initially placed within the bands will interact simultaneously, with no 
constraints due to selection rules, provided $\pi /l_s \sim 3 \Delta/2$. We shall refer to this regime as 
the regime of democratic interaction because selection rules are not operative here; under this regime, 
any three modes within the bands are coupled with the same strength. 
If $l_s \rightarrow \infty$ one recovers the matched 
selection rule $\kappa_1 = \kappa_2 + \kappa_3$, but for finite $l_s$'s any triple of modes within the 
bands are connected. The approximate dynamics 
of bands can be obtained if one assumes $\hat s(\kappa_1-\kappa_2 - \kappa_3) \sim \hat s(0)$ for 
$|\kappa_p| \leq \Delta/2$, discarding all modes outside the combs; we note that under this approximation, and 
considering the normalization choice $s(0)=1$, the first of Eqs. (\ref{eq8p5}) informs us that 
$\hat s(0) \sim 1/(3 \Delta)$. 
In this case, and moving into the discrete version of our continuum equations with 
$\kappa_{p=1,2,3} \rightarrow \kappa_m = 2 \pi m/L \equiv m \kappa_L$ (``$m$'' is an integer denoting 
the modal number), 
$d\kappa_p \hat a_p(\kappa_p) = \kappa_L \hat a_p(\kappa_p) = (2 \pi/L) \hat a_p(\kappa_p)\, \rightarrow 
\hat a_{pm}$, and $L$ as the system length, 
one arrives at the set already explored by various authors \cite{men85,men88,rob96,wil77,riz03}
 
\begin{eqnarray}
i \dot {\hat a}_{1q} = v_{g1} \kappa_q \hat a_{1q} + {1 \over 3 N_{\Delta}} \sum_{m,n} \hat a_{2m} \hat a_{3n}, 
\label{eq9} \\
i \dot {\hat a}_{2m} = v_{g2} \kappa_m \hat a_{2m} + {1 \over 3 N_{\Delta}} \sum_{q,n} \hat a_{1q}
\hat a_{3n}^\ast, 
\label{eq10} \\
i \dot {\hat a}_{3n} = v_{g3} \kappa_n \hat a_{3n} + {1 \over 3 N_{\Delta}} \sum_{q,m} \hat a_{1q}
\hat a_{2m}^\ast 
\label{eq11}.
\end{eqnarray}
To obtain set (\ref{eq9}) - (\ref{eq11}) the prefactor  $\kappa_L / (3 \Delta) = (1/3)\,(\kappa_L/\Delta)$ 
of the nonlinear terms in the discrete version is written as 
$(1/3)\,/(1/N_{\Delta})$, $N_{\Delta} \equiv \Delta/\kappa_L$ being therefore a measure of the 
number of modes composing the combs in the Fourier reciprocal space; the factor of $3$ can be absorbed into 
convenient rescalings. As mentioned, set (\ref{eq9}) - (\ref{eq11}) comprises the 
classical form of the broad band triplet interaction, where selection rules among the wave vectors 
are absent in virtue of finite size of the interaction region. Several interesting results have been 
obtained, the most prominent of which concerning the competition between the linear and nonlinear 
terms. If the linear band width terms associated with the group velocities are absent, one shows that in steady 
state the wave systems oscillates with a 
single nonlinear frequency $\Omega$. If $|\Omega|$ is larger than the largest linear frequency $v_g \Delta/2$ 
(when unnecessary, modal and comb subindexes are occasionally suppressed 
to simplify notation), a phase locking mechanism is present, preventing an 
initially coherent comb to decohere.  
In general, when a linear band width is present a time propagator $g(t)$ can be constructed 
for the total amplitude, or macroscopic field of each comb 
\begin{equation}
A_p \equiv \sum_j \hat a_{pj},
\label{eqtotal}
\end{equation}
in the form \cite{fri06}
\begin{widetext}

\begin{equation}
g \rightarrow \cases{g(t) = {1 \over \Delta}\, \int_{-\Delta/2}^{\Delta/2} -i\,{\rm e}^{i v_g\,\kappa\,t} d\kappa = 
-i\,{\sin\left({v_g \,t \, \Delta \over 2}\right) \over {v_g\, t\, \Delta \over 2}} \>\>\>{\rm (time \>\>domain)} \cr \cr
\hat g(\omega) = {\ln[(\omega - v_g \Delta/2)^2] - \ln[(\omega+v_g \Delta/2)^2] \over 2 v_g\, \Delta} - 
{i \, \pi \, {\rm Sign}(v_g \Delta/2 - \omega) + i \,\pi \,{\rm Sign}(v_g \Delta/2+\omega) 
\over 2 v_g\,\Delta} 
\>\>\>{\rm (frequency \>\>domain)}.}
\label{eq11p5}
\end{equation}

\end{widetext}
The factor $-i\,\exp(i \, v_g\,\kappa\,t)$ in the time domain expression is essentially the propagator for the microscopic mode with 
wave vector $\kappa$, and the total propagator is obtained through an integration over the whole comb. 
If in the second of Eqs. (\ref{eq11p5}) one identifies the Fourier frequency $\omega$ with the dominant nonlinear 
frequency $\Omega$, the conclusion is that a dissipative-like term arises whenever $|\Omega| < v_g\,\Delta/2$. In 
extreme nonlinear cases with $|\Omega| > v_g\,\Delta/2$, coherence is preserved. In fact, a 
relatively straightforward procedure involving expansion of $\hat g$ around $\Omega$ and 
a Fourier inversion from frequency to time domain, allows to write a coupled set for the 
the macroscopic fields \cite{fri06} which gives a good 
qualitative view of the dynamics in the democratic regime:
\begin{eqnarray}
i \dot A_1 \approx \beta_1 A_1 + {1 \over 3} A_2 A_3 
\label{mac1} \\
i \dot A_2 \approx \beta_2 A_2 + {1 \over 3} A_1 A_3^* 
\label{mac2} \\
i \dot A_3 \approx \beta_3 A_3 + {1 \over 3} A_1 A_2^* ,
\label{mac3}
\end{eqnarray}
where $\beta \rightarrow (v_g \Delta)^2/(12\,\Omega)$ if $v_g \Delta \ll \Omega$, and 
$\beta \rightarrow - i v_g \Delta$ if $v_g \Delta \gg \Omega$.
One sees that given the autonomous aspect of set (\ref{mac1}) - (\ref{mac3}) one predicts decay (shrinking of volumes 
in the corresponding phase space) if $\Delta$ becomes larger than the nonlinear frequency. 

We shall obtain $\Omega$ explicitly for some cases, but let us first dwell on the role of the width $\Delta$. 
It is a fixed quantity which corresponds to one third of the total interaction range defined by the form factor 
$\hat s(\kappa_s)$. The width $\Delta$ contains a number $N_{\Delta}=\Delta / \kappa_L$  of modes which in the past were supposed to be the 
only active modes of the wave system.  However, the traditional model set (\ref{eq9}) - (\ref{eq11}) is only an 
approximation to the full nonlinear system (\ref{eq6}) - (\ref{eq8}),
where one deliberately discards all modes outside the comb of the given width $\Delta$. The assumption looks right 
because, as mentioned, modes within the comb are expected to be more strongly and more quickly excited than modes 
outside. However, when one looks at the full set of equations there is always a nonlinear coupling which may eventually 
interlace and excite all modes, even those not initially placed inside the combs. In a real system with a band extension 
naturally much larger than the width $\Delta$, the propagator for the entire macroscopic field should be rewritten as in Eq. (\ref{eq11p5}), but 
with $\Delta$ replaced with $\Delta_T$, the latter quantity representing the total band width available to the modes. 
Thus, even if $|\Omega| > v_g\,\Delta/2$, one might still have $|\Omega| < v_g\,\Delta_T/2$, a situation where coherence 
decay might be present. Of course, if one takes $\Delta_T$ as the full band width, and $\Delta_T > \Delta$, not all modes 
will interact democratically and selection rules shall reappear. In that case, previous results must be re-evaluated. In 
particular, from the stand point of macroscopic modes, the systems ceases to be autonomous since the nonlinear terms 
can no longer be written only in terms of $A_1$, $A_2$, and $A_3$. Therefore one cannot prove or disprove that 
volumes in the phase space of the macroscopic modes are shrinking, as it happens with the approximate form given 
by Eqs. (\ref{mac1}) - (\ref{mac3}). Nevertheless a dissipative term is present and the macroscopic modes are 
likely to decay in time - this is what really happens as we show next.

\section{A more accurate view: the extended broad band interaction}

As said, the full set (\ref{eq6}) - (\ref{eq8}) is equivalent to the its counterpart spatial set 
(\ref{eq1}) - (\ref{eq3}). The connection 
is relevant because if one discards space derivatives exact solutions can be obtained. These exact solutions form the 
basis for further progress as one includes the space derivatives.

\subsection{Neglecting space derivatives}

Taking $v_{gp} \partial_x \rightarrow 0$ in the Eqs. (\ref{eq1}) - (\ref{eq3}) a stationary solution can be obtained 
in the form $A_p (x,t) = \rho_p (x) \exp \left(i \phi_p(x,t)\right)$, where $\phi_1 = -2 \rho_1(x) \, s(x) \,t$, 
$\phi_{2,3} = - \rho_1(x) \, s(x) \, t$, $\rho_{2,3} = \sqrt{2} \, \rho_1(x)$, and where $\rho_1(x)$ is an arbitrary 
$x$-dependent function; we note that in the stationary state phases depend on time, but amplitudes do not. Once 
$\rho_1(x)$ is defined, the complete solution is automatically found. And once the space time solution is found, 
Fourier transforms can be used to move into the reciprocal space. To further specify the system with initial 
conditions localized  both in the real and reciprocal spaces,  we shall make the following choice for the combs and 
the form factor $s$ in the spatial representation:
\begin{eqnarray}
\rho_1(x) = \rho_0 \, \exp\left(-x^2/l_{\rho}^2\right), 
\label{eq12} \cr
s(x/l_s) = \exp \left(-x^2/l_s^2\right).
\label{eq12pp5}
\end{eqnarray}
The comb thus defined has width $ \sim 1/l_{\rho}$ in the reciprocal space and in order that its modes interact 
democratically at least initially, we require $1/l_s \geq 3/l_{\rho}$ as explained earlier - in all numerical 
work we actually take $1/l_s = 3/l_{\rho}$. We shall also assume 
that $\Delta_T \gg \Delta$, and write for the exponential distribution $\Delta = 2\,\pi/l_{\rho}$.

Independently of the choices we make for $\rho_1(x)$ and $s(x)$ we are already in position to define coherence in 
the present case. We simply note that since 
\begin{equation}
a_p(x,t) = \sum_j \hat a_{pj} e^{i \, \kappa_j \, x},
\label{eq13}
\end{equation}
it is true that the macroscopic fields $A_p$ introduced earlier in Eq. (\ref{eqtotal}) obey simple expressions - 
we write down the one obeyed by $A_1$:
\begin{equation}
A_1 = \sum_j \hat a_{1j} = \rho_0 \exp\left(-2 \, i \,\rho_0\,t\right).
\label{eq14}
\end{equation}
From the equation above we see that the macroscopic field oscillates harmonically with frequency 
$\Omega \equiv- 2 \rho_0$ and 
with constant amplitude $\rho_0$. This is what we shall take as a coherent state: a non-decaying macroscopic mode oscillating 
with constant amplitude and constant frequency. The question that poses itself here is to determine how many 
microscopic modes actually participate in the coherent state. In other words, would be true to assume that {\it only} 
the modes inside the initially defined combs are 
active? At a first glance one might suspect the answer would be positive since those are the modes interacting more 
strongly in the system. However, we had already pointed out that due to the nonlinear cascading structure of the interaction, some 
energy may flow from low to high wave vectors; and this is what actually happens. This can be seen more formally with 
help of some tools. Keeping focus on the first comb, one first chooses a range $D$ defined by 
$-D/2 < \kappa < D/2$ and performs a partial summation over its internal modes, 
\begin{widetext}

\begin{equation}
{\cal I}_D \equiv \int_{-D/2}^{D/2} a_1(\kappa_1) d\kappa_1 = {1 \over 2 \pi} \int_{-D/2}^{D/2} \int_{-\infty}^{\infty} 
{\rm e}^{-i \kappa_1 x} \rho_1(x) \, 
\exp\left[-2 i \rho_1(x) \, s(x)\, t\right]\,dx\,d\kappa_1.
\label{eq15}
\end{equation}
\end{widetext}
For a finite band $D$, the integral over $\kappa_1$, performed firstly, 
yields a delta-like structure as a function of 
$x$, with height $D/2 \pi$ and width $2 \pi/D$. If one supposes $1/D$ small, the remaining integration over 
$x$ can be done with help of a saddle approximation near $x=0$ where the space derivative of 
fields and form factor vanish. The final result can be written in the form 
\begin{equation}
{\cal I}_D \sim {\rm e}^{-2\,i\,\rho_0\,t} \, \sqrt{\rho_0} \,
D\, {{\rm Erfi}\left[e^{3\,i \pi/4} \pi \sqrt{2\,\rho_0 \, t\, s''(0)/2}\, /D \right] \over 
\sqrt{t\,s''(0)}},
\label{eq16}
\end{equation}
where $s'' \equiv d^2s/dx^2$, where ${\rm Erfi}(\chi)$ denotes the imaginary error function as a function of argument $\chi$, 
and where we recall 
that $s(x)$ varies faster than $\rho_1(x)$. We see that all depends on the behavior of the imaginary error 
function for large and small arguments.  If $|\chi| \ll 1$, ${\rm Erfi}(\chi) \sim \chi$ and if 
$|\chi| \gg 1$, ${\rm Erfi} (\chi) \sim i$. 
One therefore concludes that 
\begin{equation}
|{\cal I}_D| \sim \cases{Constant \>\> {\rm when}\>\>t \lesssim l_s^2 \,D^2  / 2\,\pi^2\,\rho_0 \cr
{1 \over \sqrt{t}}\>\> {\rm when}\>\>t > l_s^2 \, D^2 /2\, \pi^2\,\rho_0.}
\label{eq17}
\end{equation}
In other words, given a range $D$ there is an intrinsic limiting time for coherence, 
\begin{equation}
\tau_D \equiv {D^2\,l_s^2 \over 2\,\pi^2\,\rho_0},
\label{eq17p5}
\end{equation}
where by intrinsic we understand the limiting time obtained in the absence 
of the linear frequency band width, i.e. by taking $v_g=0$. We know from our discussion regarding 
Eq. (\ref{eq8p5}) that $l_s \sim \pi / (3\,\Delta/2)$, so, the intrinsic coherence time for modes within 
the original packet would be given by $\tau_{D=\Delta} \sim 1/\rho_0$ which is relatively small since this is 
essentially the period of the nonlinear wave. Our conclusion is that the initial packet can be hardly 
called a coherent structure even in the absence of the frequency band width. The collection of modes 
that could be seen as a coherent structure is anyone where $D \gg \Delta$. In that case it is still 
true 
that decay will be present, but for all practical purposes $\tau_D$ would be so much larger than the period of 
the nonlinear wave that a physical setting or equipment resolving modes up to $\kappa \sim D$ would perceive the 
wave system as coherent. 

A second important time scale has to be defined for the wave system. It is the time scale of excitation of individual 
modes in the reciprocal space. Looking again at the first comb - reasonings are similar for the 
other two, we first recall the expression 
$A_1(x,t) = \rho_1(x) \exp\left(i \phi_1(x,t)\right)$ for the steady state field. $\rho_1(x)$ is constant in time and 
the phase $\phi_1(x) = -2 \rho_1(x) \, s(x) \, t$ depends both on the spatial coordinate and time. If one evaluates 
the phase gradient $\partial \phi_1 / \partial x$ and look at the maximum of this quantity as the largest wave 
vector involved in 
the dynamics, one derives the relation 
\begin{equation}
|\kappa_{max}| \sim (2 \rho_0 /l_s)\, t,
\label{eq18}
\end{equation}
which shows that the packet 
spreads over the reciprocal space at a rate $\pm2 \rho_0 /l_s$. The time for excitation of any particular wave vector 
$\kappa_{max}$ is thus $\tau_{exc}(\kappa_{max}) = \kappa_{max} l_s/2 \rho_0$. If we take $\kappa_{max} = D/2$, we 
see that for the typical case $D \, l_s \gg 1$, $\tau_D \gg \tau_{exc}(\kappa_{max}=D/2)$, which means that in the absence of 
linear frequency band widths, coherence time of a packet of range $D$ is in general much longer than the time 
required to activate the modes at the borders of the packet.

\begin{figure}
\includegraphics[scale=0.4]{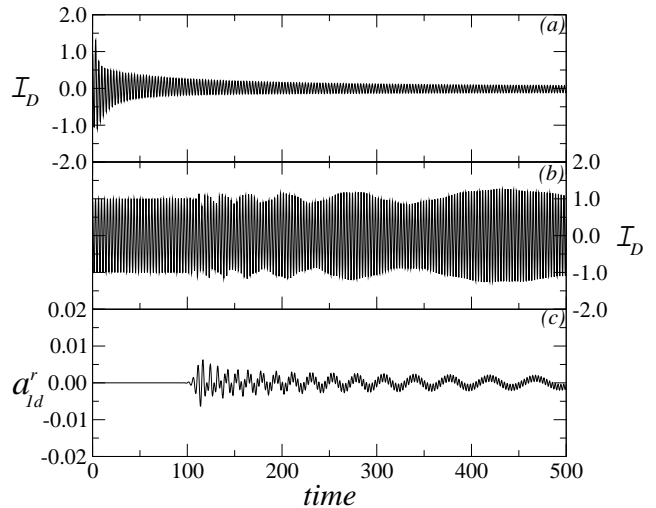}
\caption{${\cal I}_D$ as a function of time for $D\,l_s=2\,\pi \times 2^2$ in (a) and for 
$D\,l_s=2\,\pi \times 2^6$ 
in(b). In (c) we show the times series for the real part of the borderline mode with wave vector 
$\kappa = D/2$, $l_s = 2^9$, $l_{\rho} = 3 \> l_s$. The group velocity is zero for all waves and 
all quantities are dimensionless.
\label{fig1.eps}}
\end{figure}

In Fig. \ref{fig1.eps} we display the contrasting behaviors for ${\cal I}_D$ in the cases 
$D \, l_s \sim 1$ and $D \, l_s \gg 1$. In the simulations we integrate set (\ref{eq1}) - (\ref{eq3}) with a 
pseudo-spectral method, using a grid of length $L=N=2^{15}$, $N$ denoting the number of nodes which 
for scaling simplicity is equal to the length. In all numerical analysis we use $\rho_0=1$; the choice is not 
restrictive because field scales can always be absorbed in space and time. 
Considering $l_s = 2^9$ and $l_{\rho} = 3\, l_s$, panel (a) displays the case $D=2 \pi/L \times 2^8$, 
for which $D \, l_s = 2\,\pi \times 2^2$, and $\tau_D \sim 32$, while in panel (b) 
$D \, l_s = 2\,\pi \times 2^6$ for which $\tau_D \sim 8200$. 
Panel (a) reveals a fast decay, but coherence is far more persistent 
in panel (b). It is noticeable that in panel (b) function ${\cal I}_D$, although initially laminar, develops slight 
modulations after a very sharp instant along the time axis. This very sharp instant corresponds to $t=\tau_{exc}(D/2)$. 
Indeed, the excitation time reads $\tau_{exc}(D/2)=100.5$ in this particular instance. This is confirmed in panel (c) 
where, for the same parameters of panel (b), we show the time evolution of the 
real part of mode with wave vector $D/2$. 
The superscript ``$r$'' means ``real part'' and the submodal index $d$ reads 
$d \equiv (D/2)/(2 \pi/L) = 2^{11}$ in this case, as defined in the context of the discrete equations, 
Eqs. (\ref{eq9}) - (\ref{eq11}). 
We emphasize that as mode $\kappa=D/2$ is excited, 
coherence of the packet $D$, although undergoing a modulational process, does not decay.

Of course, the presence of a band width for the linear frequencies may change the entire picture, and 
this is the subject of the next section.

\subsection{The effects of space derivatives and the associated linear frequency band width}

Since the full nonlinear system is not autonomous from the perspective of macroscopic modes, one cannot 
make very formal predictions about coherence decay due to the frequency band widths, like we did in the 
approximations leading to Eqs. (\ref{mac1}) - (\ref{mac3}). 
However, some estimates can still be made. 

\begin{figure}
\includegraphics[scale=0.4]{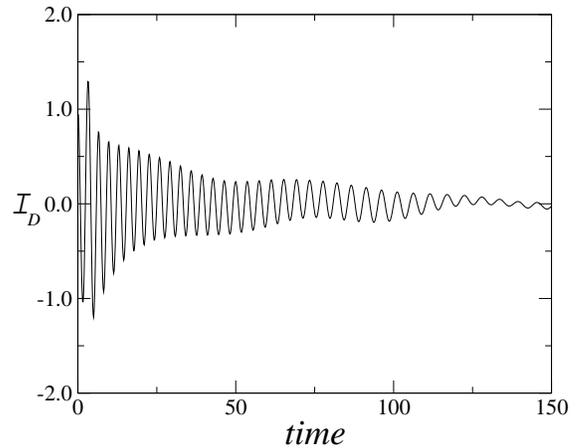}
\caption{Coherence decay due to the resonant effect. Parameters are those of panel (b) of Fig. \ref{fig1.eps}, 
with exception of $v_{g1}$ which here reads $v_{g1}=1/(2^9 \, \kappa_L)$, defining a resonant time $\tau_r \approx 50$.
\label{fig2.eps}}
\end{figure}

Let us consider our expression (\ref{eq18}) for the maximal wave vector involved into the dynamics. 
When $\kappa_{max}$ reaches the value $\kappa_r$ corresponding to the resonant frequency, 
$v_{g1} \kappa_r \equiv 2\,\rho_0$, coherence is expected to be lost, but now 
due to resonant effects. Under this circumstance, the largest excited linear frequency would 
become comparable to the nonlinear triplet frequency $\Omega$, and coherent nonlinear effects would be 
no longer dominant.  
The time to attain resonance, let us call it $\tau_r \equiv \tau_{exc}(\kappa_r)$, can be obtained as one uses 
$\kappa_{max}=\kappa_r$ in 
Eq. (\ref{eq18}): 
\begin{equation}
\tau_r \equiv l_s/v_{g1}.
\label{eq19}
\end{equation} 
A given collection of modes of range $D$ will remain coherent as long as 
$t < \tau_{coh} \equiv \min \{\tau_D,\tau_r\}$.  To illustrate this point, let us take the case 
analyzed in the panel (b) of Fig. \ref{fig1.eps}. In that 
case $\tau_D$ is large and we do not expect to see coherence decay soon if the linear frequencies are 
absent. But now let us add a frequency band width with $v_{g1}$ chosen such that a given mode 
of the spectrum becomes resonant with $\tau_r < \tau_D$; we achieve this requirement 
with $v_{g1} = 1/(\kappa_l\,2^9)$ which yields $\tau_r \approx 50 < \tau_D$. For completeness we take $v_{g1,2}=0$ which 
corresponds to one wave moving relatively to the other two. The setting would be of relevance to 
Brillouin scattering, for instance, where two electromagnetic waves with the same group velocity interact with a 
slower ion wave; we would be examining the process in the frame where the electromagnetic wave is 
stationary. The resulting dynamics is then displayed 
in Fig. \ref{fig2.eps}, where one clearly sees a fast decay whereas for $v_g=0$ one sees persistent 
coherence as previously shown in Fig. \ref{fig1.eps}(b).

Expressions (\ref{eq17p5}) and (\ref{eq19}) therefore provide us with a simple tool 
to make estimates on the circumstances allowing coherence to be seen in the nonlinear triplet system. 
Once one has defined an extended comb distributed over a range $-D/2 < \kappa < D/2$ of wave vectors 
with $D/\Delta \gg 1$, and once one knows the group velocity $v_g$ for this particular class of 
wave, the coherence time can be obtained. 

\section{Final remarks}

In this paper we developed a technique to investigate 
coherence in nonlinear triplets, when the available band of modes is much larger than that of the initial 
combs. If modes remain restricted to their initial combs, the series of approximations outlined in 
\S 2 allows to describe the system as an interaction of macroscopic modes. 
In the presently studied case, one cannot resort to these approximations because initially low 
amplitude, idle modes 
outside the initial range will be gradually excited at a rate $\sim 2\,\rho_0/l_s$, whenever the whole available 
band is larger than $\Delta$. Coherence in this, perhaps, more realistic case is a little more involved 
subject to define. One first defines the range $D$ of interest. The range has an intrinsic coherence 
time $\tau_D$ defined in the absence of any frequency band width, i.e., for $v_g=0$: 
$\tau_D = D^2 \, l_s^2/ (2\,\pi^2 \rho_0)$. 
$\tau_D$ is the largest coherence time of a collection of modes contained within the limits 
$-D/2 < \kappa < D/2$. Then, once the range $D$ is defined, one has to look at the excitation time $\tau_r$ of 
the resonant mode, which does not necessarily belongs to the range $D$; we found that 
$\tau_r = l_s/v_g$. Gathering together both time scales, the final 
conclusion is that the coherence time $\tau_{coh}$ satisfies 
$\tau_{coh} = \min\, \{\tau_D,\tau_r\}$. We have also observed and stressed that coherence gains some 
substantial meaning only when several nonlinear oscillations occur prior to $\tau_D$. Since in our 
normalized variables the period of the nonlinear oscillation is $\sim 1/\rho_0$, one concludes that 
the dynamics resembles a nonlinear phase locking process only when 
$D \gg \Delta$ and $v_g \, \Delta \ll \rho_0$. 

Let us connect our results with those of previous works. Our macroscopic model does not look into fine microscopic 
scales of size, say $l_{mic}$, where discrete effects become relevant. Therefore an upper limit $D_{max} \sim 1/l_{mic}$ does exist beyond which mode dynamics is 
naturally attenuated by microscopic effects. One can however imagine that modes with wave vectors 
$|\kappa| > D_{max}/2$ are initially small and heavily damped; if this is true they will be minimally excited 
during the dynamics. Under these circumstance the condition on $\tau_D$ for an inaccessible 
$D > D_{max}$ ceases to exist (since $\tau_D \rightarrow \infty$ in this case) and we are left only with the 
condition on the group velocity and linear band width, which is similar to what is discussed in previous 
investigations. For $D<D_{max}$ $\tau_D$ is finite and physically relevant.

\bigskip

We acknowledge support by CNPq, Brasil, and by the
AFOSR, USA, under the grant N$^{\underline o}$ FA9550-06-1-0345. 
We also thank illuminating discussions with R. Pakter and A. Serbeto.

%
%

%
%
\end{document}